\documentclass[%
 aip,
 amsmath,amssymb,
 reprint,%
]{revtex4-1}

\usepackage{graphicx}% Include figure files
\usepackage{dcolumn}% Align table columns on decimal point
\usepackage{bm}% bold math
\usepackage[utf8]{inputenc}
\usepackage[T1]{fontenc}
\usepackage{mathptmx}
\usepackage{etoolbox}
\usepackage{chemformula}
\usepackage{float}
\usepackage{xcolor}
\makeatletter
\def\@email#1#2{%
 \endgroup
 \patchcmd{\titleblock@produce}
  {\frontmatter@RRAPformat}
  {\frontmatter@RRAPformat{\produce@RRAP{*Authors to whom correspondence should be addressed: \href{mailto:#2}{#2; 
  hyhwang@stanford.edu}}}\frontmatter@RRAPformat}
  {}{}
}%
\makeatother
\begin{document}

\preprint{AIP/123-QED}

\title[Fracture and Fatigue of Thin Crystalline \ch{SrTiO3} Membranes ]{Fracture and Fatigue of Thin Crystalline \ch{SrTiO3} Membranes }
% Force line breaks with \\
\author{Varun Harbola* }
\email{varunh@stanford.edu}
\affiliation{Department of Physics, Stanford University, Stanford, California 94305, USA}
\affiliation{Stanford Institute for Materials and Energy Sciences, SLAC National Accelerator Laboratory,
Menlo Park, California 94025, USA} 


\author{Ruijuan Xu}%
\affiliation{Stanford Institute for Materials and Energy Sciences, SLAC National Accelerator Laboratory,
Menlo Park, California 94025, USA} 
\affiliation{Department of Applied Physics, Stanford University, Stanford, California 94305, USA 
%\\This line break forced with \textbackslash\textbackslash
}%

\author{Samuel Crossley}
\affiliation{Stanford Institute for Materials and Energy Sciences, SLAC National Accelerator Laboratory,
Menlo Park, California 94025, USA}
\affiliation{%
Department of Applied Physics, Stanford University, Stanford, California 94305, USA %\\This line break forced% with \\
}%

\author{Prastuti Singh}
\affiliation{Stanford Institute for Materials and Energy Sciences, SLAC National Accelerator Laboratory,
Menlo Park, California 94025, USA}
\affiliation{%
Department of Applied Physics, Stanford University, Stanford, California 94305, USA %\\This line break forced% with \\
}%

\author{Harold Y. Hwang*}

\affiliation{Stanford Institute for Materials and Energy Sciences, SLAC National Accelerator Laboratory,
Menlo Park, California 94025, USA}
\affiliation{%
Department of Applied Physics, Stanford University, Stanford, California 94305, USA %\\This line break forced% with \\
}%
%\email{hyhwang@stanford.edu}

\date{\today}% It is always \today, today,

\begin{abstract}
The increasing availability of a variety of two-dimensional materials has generated enormous growth in the field of nanoengineering and
nanomechanics. Recent developments in thin film synthesis have enabled the fabrication of freestanding functional oxide membranes that can
be readily incorporated in nanomechanical devices. While many oxides are extremely brittle in bulk, recent studies have shown that, in thin
membrane form, they can be much more robust to fracture as compared to their bulk counterparts. Here, we investigate the ultimate tensile
strength of \ch{SrTiO3} membranes by probing freestanding \ch{SrTiO3} drumheads using an atomic force microscope.  
We demonstrate that \ch{SrTiO3}
membranes can withstand an elastic \textcolor{black}{deformation with an average strain} of \textasciitilde 6\% in the sub-20 nm thickness regime, which is more than an order of magnitude beyond the
bulk limit. We also show that these membranes are highly resilient upon a high cycle fatigue test, \textcolor{black}{surviving up to a billion cycles of force modulation at 85\% of their fracture strain}, demonstrating their high potential for
use in nanomechanical applications.
\end{abstract}

\maketitle

Since the discovery of graphene, two-dimensional (2D) materials have attracted a great deal of attention for not only their electronic
properties, but also their mechanical characteristics. These materials exhibit both a large elastic modulus and high tensile
strength\cite{Bertolazzi2011, lee:2008} making them attractive for nano-electromechanical applications. Traditionally, oxide materials can
also be grown to be extremely thin as films and heterostructures, and have an exciting array of physical properties, from dielectrics to
magnetism to superconductivity. However, thin films grown on a rigid substrate are not ideal for nanomechanical applications due to the
clamping effect from the substrate. Recent technique developments allow for these thin films to be grown epitaxially and then lifted off
and transferred, such that they form freestanding structures\cite{Gan1998,Paskiewicz2016,Bakaul2016,Lu:2016,Ji2019,Kum2020,pesquera:2020}. This has
enabled studies characterizing the mechanical properties and the effects of different deformations in freestanding oxide membrane
structures\cite{harbola:2021,Davidovikj2020}. Furthermore, strain mapping using transmission electron microscopy (TEM) has shown these membranes can withstand up to 10\% strain locally\cite{Dong2019,Peng2020},  undergoing extreme deformations without breaking\cite{Elangovan2020}. The proposed mechanisms for high strain sustenance by these membranes were low numbers of flaws in smaller samples, continuous dipole rotation \textcolor{black}{in ferroelectrics }during deformation that avoids sharp domain-switching-driven failure\cite{Dong2019}, and proximity to a strain-induced phase transition\cite{Peng2020}.

In this work, we consider the fracture properties of a canonical perovskite membrane, \ch{SrTiO3}. \ch{SrTiO3} is a high\textcolor{black}{-K} dielectric insulator at room
temperature with a relative permittivity of 300 in bulk. Moreover, it is a transparent oxide with a 3.2 eV bandgap and at low temperatures
exhibits dilute superconductivity upon doping\cite{Schooley1964}. \ch{SrTiO3} at low temperature shows a multitude of structural transitions\cite{Scott1997} and even though
its permittivity greatly increases at low temperatures, \ch{SrTiO3} never achieves a ferroelectric state due to quantum fluctuations\cite{Muller1979}. However,
upon straining, both on substrate\cite{Haeni2004,Biegalski2009} and in freestanding form\cite{Xu2020}, ferroelectricity has been demonstrated in \ch{SrTiO3}. Nano resonators made
from \ch{SrTiO3} membranes have also been shown to have high Q values and low mechanical losses at low temperatures\cite{Davidovikj2020}. \textcolor{black}{\ch{SrTiO3} cantilevers formed using an under etch method have also shown promise demonstrating their capability to be electromechanically actuated at the microscale\cite{Biasotti_2013}.} All these properties make
\ch{SrTiO3} a promising material for nanomechanical applications, making it important to study the robustness of this material under strain both
in terms of fracture and fatigue.

\begin{figure*}[t]
    \centering
    \scalebox{0.083}{\includegraphics{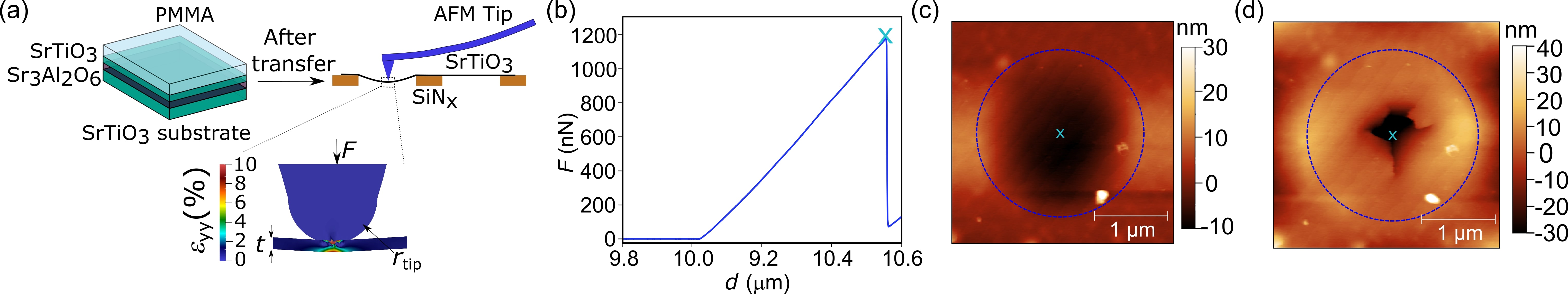}}
    \caption{(a) A schematic representation of the experiment \textcolor{black}{where the \ch{SrTiO3} membrane is grown on an epitaxial strontium aluminate buffer layer  and transferred to a porous silicon nitride grid. } An AFM probe of radius $r_\textrm{tip}$ forces the membrane of thickness $t$ to its fracture with a force $F$. A finite element strain map \textcolor{black}{around the tip-membrane contact region} shows that the maximal strains are concentrated under the tip. (b) A force-displacement ($F$-$d$) curve showing the force response and identification of the force at which the membrane fractures. Panels (c) and (d) show AFM topography images of a \ch{SrTiO3} drumhead before and after fracture, respectively.}
    \label{fig: my_label}
\end{figure*}

To quantitatively establish the fracture of a material, the essential requirement is knowing the stress at which the material fails. To
measure the failure point of bulk materials, generally, a sample of a known cross-section is stretched to its breaking point, and the force
at which it breaks divided by the cross-sectional area defines the fracture stress. Such a direct measurement is not always feasible at the
nanoscale, and a standard approach is to use a freestanding geometry of the nanomaterial, which is then forced with a calibrated nano-probe
until its fracture\cite{Bertolazzi2011,lee:2008,Kaplan-Ashiri2006,Yu2000}. In this work, we study the fracture of \ch{SrTiO3} membranes via atomic force microscopy (AFM). \ch{SrTiO3} membranes
were grown using pulsed laser deposition (PLD) on a water-soluble and epitaxial buffer layer \textcolor{black}{of 8 nm thick \ch{Sr3Al2O6} on a \ch{SrTiO3} substrate.} \textcolor{black}{The membranes is spun coated with PMMA and then} lifted off \textcolor{black}{using water} and transferred
onto a porous silicon nitride grid\cite{harbola:2021}. These membranes have been shown to retain high levels of crystallinity down to thicknesses of 2 nm\cite{Hong2017}.
This transfer forms freestanding drumheads of \ch{SrTiO3} on the nitride membranes. These drumheads can then be probed using an AFM tip until
they rupture to study the fracture mechanics of thin \ch{SrTiO3} membranes (Fig. 1(a)). Recently, a non-monotonic variation in Young’s modulus
with thickness was observed in \ch{SrTiO3} membranes, as a consequence of strain gradient elasticity. Therefore, we study three different
characteristic thicknesses for fracture in the sub-15 nm regime, where a softening of Young’s modulus was observed with increasing
thickness\cite{harbola:2021}. 

To measure the force at which freestanding membranes break, the deflection of the membrane is registered using a photodiode. The spring
constant of the tip cantilever is calibrated using the thermal method\cite{Cook2006}. 
The force at which the membrane fractures on tip impact can be
obtained through a force-$d$ curve (Fig. 1(b)), where $d$ is the distance traveled of the z-direction piezo of the AFM. This can be used to
quantify the average 2D stress under the tip using\cite{Bhatia1968}

\begin{equation}
\sigma_mt =  \left(\frac{FEt}{4\pi r_\textrm{{tip}}}\right)^\frac{1}{2}
\end{equation}
where $F$ is the maximum force sustained by the drumhead before breaking, $E$ is the Young’s modulus, 
$t$ is the thickness of the membrane, $r_\textrm{tip}$ is the radius of curvature of the AFM probe tip, 
and $\sigma_m$ is the average stress that is sustained by the membrane under the tip (Fig. 1(a); see supplementary material for more discussion on strain distribution).
The tips are imaged using a scanning electron microscope (SEM) to estimate the tip radius, found to be 14 $\pm$ 3 
nm across different tips, which is comparable to manufacturer specifications. First, we scan the drumhead in tapping mode for topography
and position the tip at the center of the drumhead (Fig. 1(c)). Once the tip is in position, we use the $d$ position of the piezo as the
trigger to increase the force on the membrane until it breaks. This measurement is quasi-static in nature and the tip is always in contact
with the membrane while forcing it, so the force applied by the tip is the force felt by the membrane to the point of fracture. A
topography image of the broken drumhead clearly shows the rupture at the center of the drumhead. During the forcing process, the
repeatability of the force traces indicates that there is no slippage or plastic deformation. The fracture is brittle, showing no yielding
of the material before the circular membrane ruptures.

\begin{figure}[h]
    \centering
    \scalebox{0.12}{\includegraphics{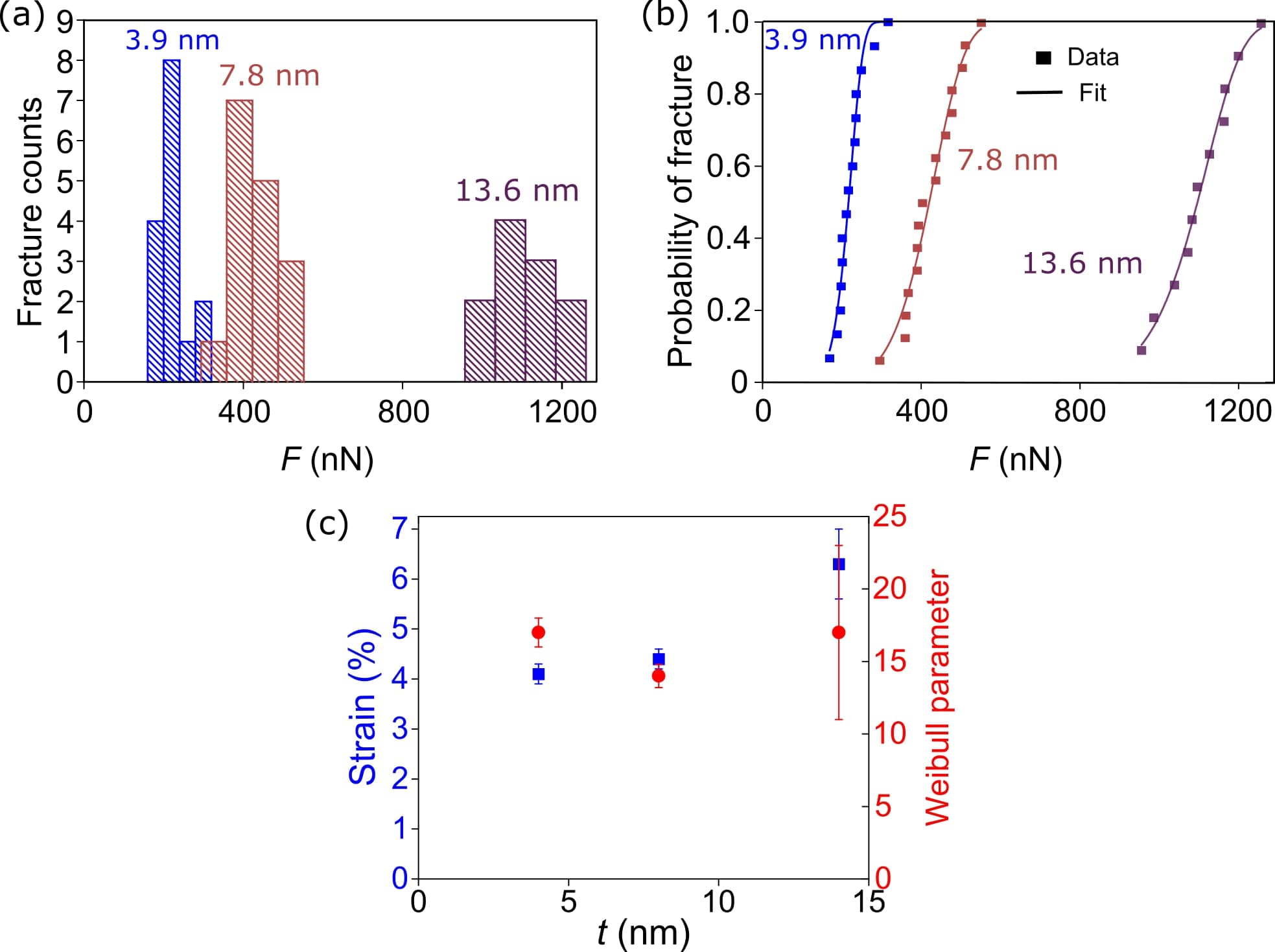}}
    \caption{(a) A histogram plot of fracture statistics with respect to the maximum force sustained by the membrane before fracture for three different thicknesses. (b) Same data as (a) but plotted as a cumulative probability of fracture. The solid line is a 2-parameter Weibull fit to the experimental data. (c) Plot of the statistical fracture strain and the Weibull parameter $m$ as a function of thickness obtained through Weibull analysis of the data. Error bars include errors from spring constant calibration of the AFM cantilever and errors in the radius estimation of the tip. For the 13.6 nm sample, a total of three different membranes were tested so that error includes the standard error from 3 samples.}
    \label{fig: my_label}
\end{figure}

Fracture of a material is a statistical process which is governed by a variety of factors such as the types of defects, their density, and the flaw size distribution. Real materials will always have a distribution of stresses at which various samples will fail, even when they are prepared identically. The determining factor for sample fracture is the extremal size distribution of flaws in the effective volume where the sample is experiencing stresses. A two-parameter Weibull distribution appropriately describes the cumulative probability of fracture for brittle materials as a function of stress $(P(\sigma))$ \cite{lee:2008,Quinn2010}. It is given as

\begin{equation}
    P(\sigma)=1-e^{-\left(\frac{\sigma}{\sigma_0}\right)^m}
\end{equation}
where $\sigma_0$ is the characteristic stress of fracture and $m$ is the Weibull shape parameter and describes the sharpness of this
distribution. A low $m$ value is indicative of a wide distribution of failure stress, which implies that a wide distribution of defects is
responsible for failure. On the other hand, a higher $m$ value indicates either an insensitivity to the presence of defects, or a very
narrow distribution of defects which are responsible for material failure\cite{lee:2008}. The higher the $m$ value, the more predictable
the failure of a material becomes. Moreover, as Weibull statistics are rooted in extremal flaw size distribution, Weibull analysis of a
sample failure also allows for scalability when the size of the sample is changed\cite{Quinn2010}. Note that the stress is normalized by $\sigma_0$, such that
Weibull distributions having the same $m$ values will have a wider variance of breaking stress for higher $\sigma_0$. 

The statistical distribution of \ch{SrTiO3} drumheads fracture (Fig. 2(a)) shows a clear peaked distribution as a function of force for all
three different thicknesses studied. These distributions can be changed into cumulative fracture distributions and can be analyzed using
the Weibull distribution curve. Fig. 2(b) shows that the fracture of \ch{SrTiO3} drumheads is well described by the two parameter Weibull fit.
The analysis indicates that one type of flaw is responsible for the failure of these drumheads\cite{Quinn2010}. Using Eq. (1), the $m$ value obtained
through the fit as a function of force can be mapped to a corresponding $m$ value for stress by multiplying by a factor of two. 
The stress   can also be converted to the maximum strain sustained by the film using the Young’s modulus of \ch{SrTiO3} for stretching which has
been measured previously\cite{harbola:2021} for these thicknesses. We find that the films can sustain an average strain of 4-6 \% before fracture and the $m$
value is close to 16 for thin \ch{SrTiO3} (Fig. 2(c)). Using finite element analysis calculations, we also observe that the maximum local strain
sustained by these \ch{SrTiO3} membranes is reasonably consistent with that observed via TEM\cite{Dong2019,Peng2020} for freestanding oxide membranes (supplementary material). 

\begin{figure}[h]
    \centering
    \scalebox{0.15}{\includegraphics{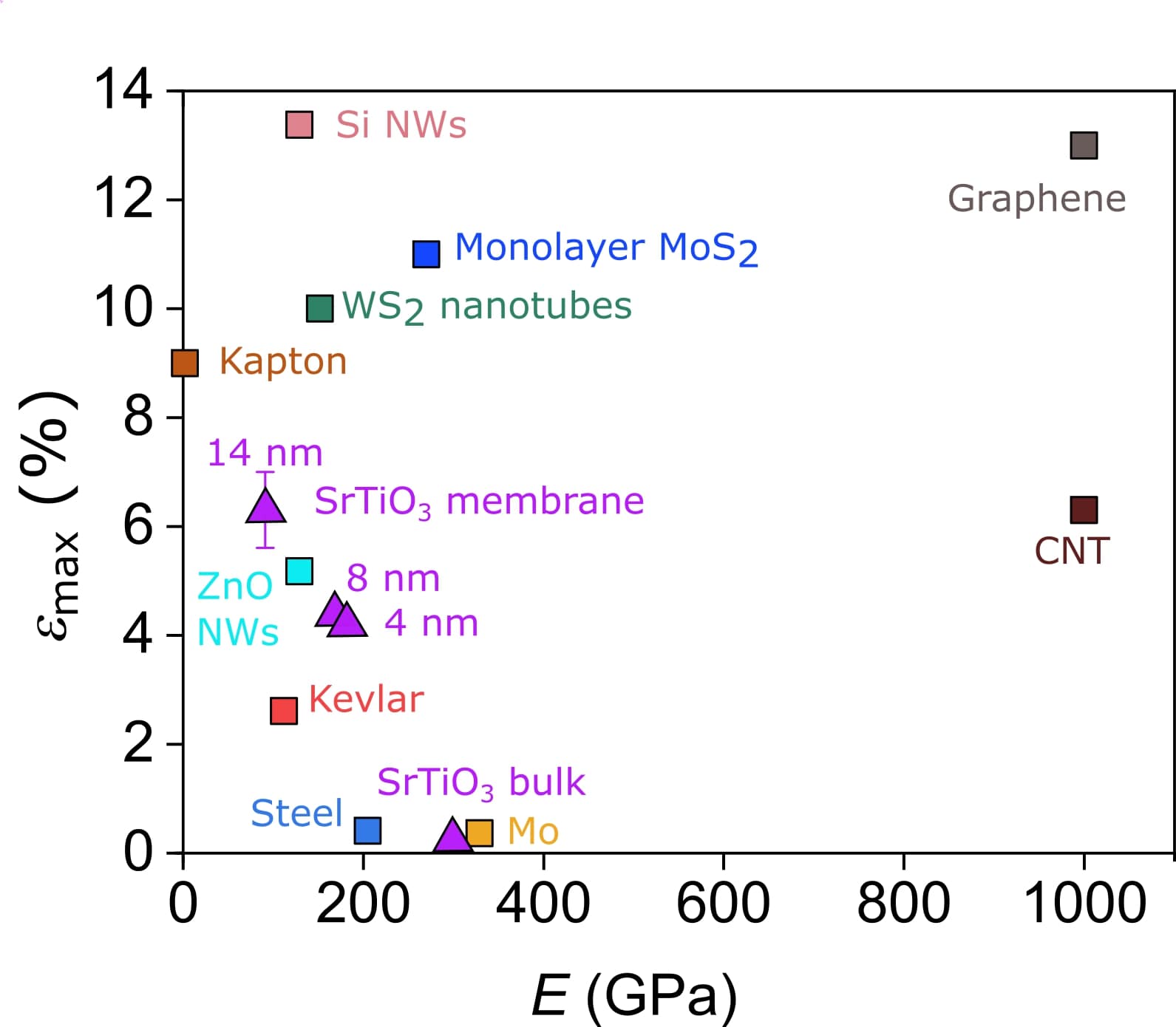}}
    \caption{(a) A scatter plot of different materials covering a range of both
    nanomaterials\cite{Bertolazzi2011,lee:2008,Kaplan-Ashiri2006,Yu2000,Roy2017,Zhang2016} and bulk\cite{Gumbsch2001,Shinno1988,Bunsell1975,Rasmussen2003,dupont1} representing the
    maximum tensile strain ($\varepsilon_\textrm{max}$) these materials can sustain before fracture, plotted against their respective elastic moduli. The \ch{SrTiO3} membranes have three different moduli for the three different thickness and are therefore plotted separately\cite{harbola:2021}.}
    \label{fig: my_label}
\end{figure}

Let us place these results within the context and understanding of other materials. \ch{SrTiO3} in bulk form is extremely brittle and can only
sustain a fraction of a percent of strain at room temperature before breaking. However, in the thin membrane form, it is able to sustain
more than an order of magnitude higher strain than in bulk. Fig. 3 shows the maximum tensile strain before fracture for a variety of
materials plotted against their elastic modulus. Given the variation in the elastic modulus of \ch{SrTiO3} membranes as a function of
thickness\cite{harbola:2021}, each thickness of \ch{SrTiO3} membranes measured in this study has been plotted separately in Fig. 3. The strain that thin \ch{SrTiO3}
membranes can withstand is similar to that displayed by carbon nanotubes\cite{Yu2000} and ZnO nanowires\cite{Roy2017}. 
\begin{table}[h]

\begin{ruledtabular}
\begin{tabular}{lcr}
Material & Weibull Parameter $m$\\
\hline

\ch{WS2} nanotubes\cite{Pugno2007} &  2.9 \\
Carbon nanofibers \cite{Pugno2007} &  3.8\\
Carbon MWNT \cite{Pugno2007}  &  2.7\\
ZnO nanowires \cite{Roy2017} & 3.9\\
Graphene monolayer \cite{lee:2008} & 32\\
\ch{SrTiO3} membrane & 16 \\
Polysilicon (MEMS) \cite{Jadaan2003} & 5-30\\
Single crystal Si \cite{Jadaan2003} & 3-60\\

\end{tabular}
\end{ruledtabular}
\caption{\label{tab:table1}Comparison of fracture robustness among materials. This table lists a variety of nanomaterials which have been tested for their robustness of fracture with respect to the stress at which they fracture, quantified by their Weibull shape parameter $m$. }
\end{table}
Also notable is that these oxide membranes can bear up to about half the strain of that in graphene\cite{lee:2008}, which is the strongest material known thus far and can sustain up to 13\% strain before breaking. In terms of the $m$ parameter (Table 1), \ch{SrTiO3} membranes compare extremely well with other nanomaterials. The predictability of failure for \ch{SrTiO3} membranes is much higher than a variety of nanofibers\cite{Pugno2007} and nanotubes\cite{Yu2000} and is comparable to single crystal and polycrystalline silicon micro-electromechanical systems (MEMS)\cite{Jadaan2003}. Furthermore, we can use this $m$ value to predict failure of thin membranes in different experiments by using volume scaling via:\cite{Quinn2010}
\begin{equation}
    \frac{\sigma_{m1}}{\sigma_{m2}}=\left( \frac{V_2}{V_1} \right)^{\frac{1}{m}}
\end{equation}
where $\sigma_{m1}$ and $\sigma_{m2}$  are the maximal stresses that can be sustained in experiment 1 and experiment 2 and $V_1$ and $V_2$
are the effective volumes over which the stresses are being imparted in those experiments, respectively. 
We can estimate the effective volume for our indentation experiment using finite element simulations (Fig. 1(a) and supplementary 
material). This volume is approximately a 10 nm radius region under the tip across the thickness of the membrane. Using this estimated
volume, we can predict the breaking strain for a separate experiment that was performed for large area \ch{SrTiO3} strained membranes on Kapton,
in which case the whole volume of the membrane experienced the applied stress during the experiment\cite{Xu2020}. Using Eq. (3), we can estimate
the maximum strain for the Kapton experiment. Most membranes failed at around 2\% upon stretching on Kapton, which is consistent with the average estimate of 1.8\% from Weibull scaling of our AFM measurement. The highest estimate of fracture strain from the 14 nm thick membranes is 2.7\%, which is close to the maximum strain of 2.5\% observed\cite{Xu2020}.

\begin{figure}[h]
    \centering
    \scalebox{0.12}{\includegraphics{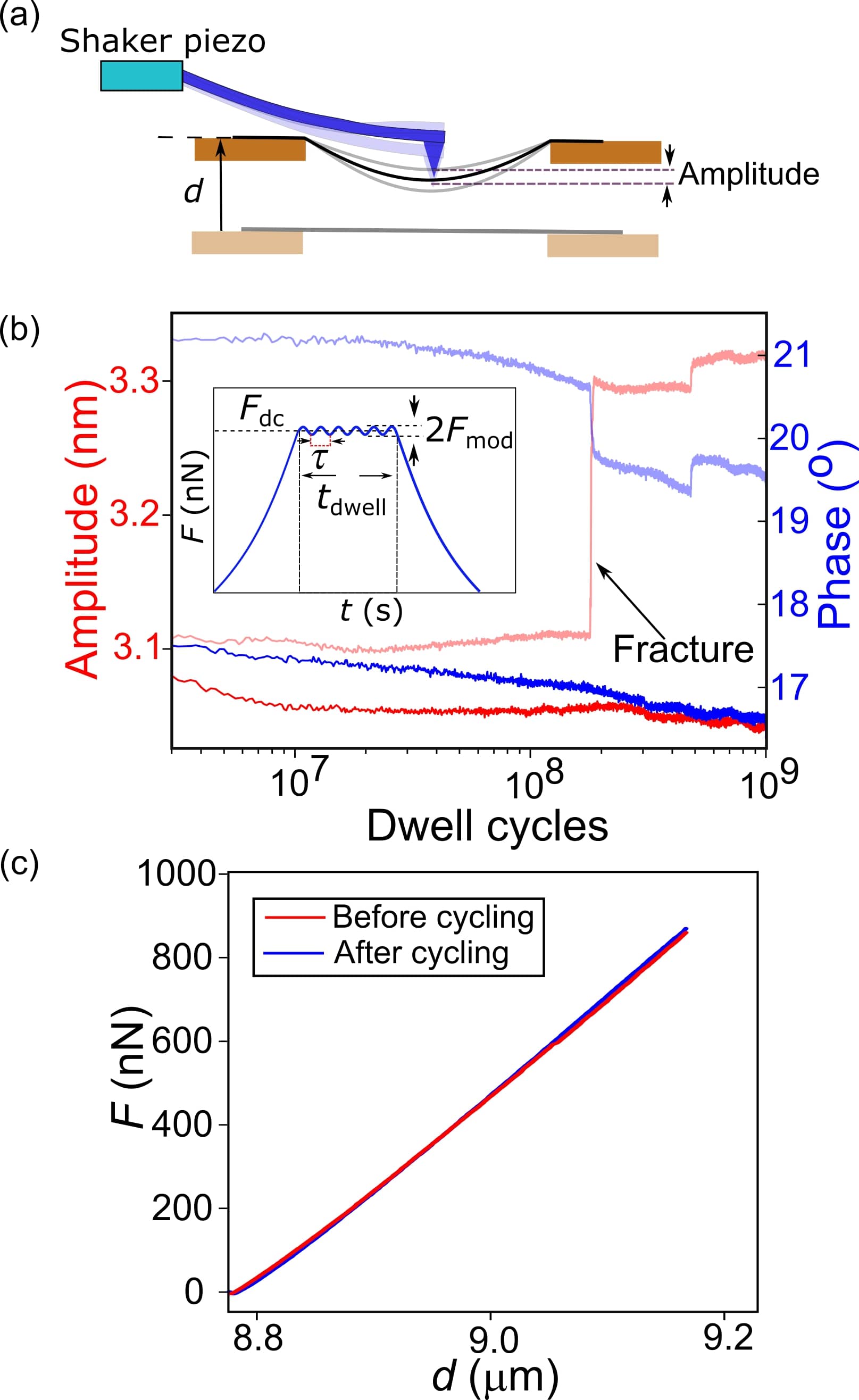}}
    \caption{(a) Schematic representation of a high cycle fatigue test of a freestanding \ch{SrTiO3} membrane. \textcolor{black}{The $z$-direction piezo moves a distance $d$ to force the membrane from its initial state to a high strain state and then} the modulation is provided using the shaker piezo of the tip usually used for tapping mode scans. \textcolor{black}{This tip modulation probes fatigue of the membrane. } (b) The amplitude and phase response of freestanding membranes (13.6 nm thick) upon a high cycle fatigue test over a billion cycles. Faded color is the response of the only membrane that fractured out of the fourteen tested. Inset shows the experimental procedure of a high cycle fatigue test whereupon the membrane is forced with a constant force $F_\textrm{{dc}}$, which corresponded to ~85\% of the breaking stress on these membranes, and then cycled with a modulation force $F_\textrm{{mod}}$ over a set number of cycles. (c) The force response of a membrane before and after the fatigue test showing almost identical response, demonstrating a lack of fatigue degradation after one billion cycles.}
    \label{fig: my_label}
\end{figure}

Thus far we have demonstrated that \ch{SrTiO3} membranes show more than an order of magnitude higher stress sustenance compared to the bulk,
with a high predictability of failure. We next studied their fatigue properties, which have not yet been investigated for this class of
complex oxide membranes. Fatigue occurs due to bond reconfigurations and plastic deformations near a defect upon cyclical force loading.
Since ionic bonds are not easily reconfigurable and they fail in a brittle manner, oxides are one of the least prone materials to fatigue.
Only recently has attention started turning to fatigue properties of nanomaterials\cite{Li2014,Cui2020}, and with the growing prevalence of diverse
nanomaterials for nano-electromechanical purposes, such measurements become increasingly important to predict the lifetime of
nanocomponents. To study the fatigue properties of \ch{SrTiO3} membranes, we used the AFM in dwell mode to conduct the experiment of high cycle
fatigue (Fig. 4(a)). We used the $d$ trigger on the AFM to force the membrane to the required force, and then modulated the tip at that $d$
value with a frequency of 2 MHz for 500 seconds, to force the membrane through a billion cycles (Fig. 4(b)) with a modulation force of 10
nN. Out of the 14 drumheads tested for fatigue at over 85\% of the characteristic fracture strain, only one showed fatigue failure at close
to 200 million cycles. Moreover, upon testing the repeatability of elastic response after cyclic loading, no sign of fatigue yield was
found (Fig. 4(c)). This result indicates that \ch{SrTiO3} membranes are comparable to ZnO\cite{Li2014} and graphene\cite{Cui2020} as far as their fatigue behavior is concerned,
over a large number of cycles under high static stress.

To conclude, we performed a detailed statistical analysis of fracture of freestanding \ch{SrTiO3} membranes and demonstrated that their
nanomechanical fracture is robust and well explained through Weibull statistics. \ch{SrTiO3} membranes show more than one order of
magnitude enhancement in their strain sustenance as compared to the bulk upon local loading. Furthermore, through two-parameter Weibull
analysis we showed that the predictability of failure for \ch{SrTiO3} is significantly higher than for many other nanomaterials, and on par
with silicon and graphene. We have also shown excellent fatigue resilience of these membranes under high stress and over a billion cycles.
These findings add to the growing body of evidence that thin freestanding oxide membranes are an extremely viable class of materials
for nano-electromechanical applications.

\begin{acknowledgements}

We thank Wendy Gu for discussions. This work was supported by the U.S. Department of Energy, Office of Basic Energy Sciences, Division of Materials Sciences and Engineering, under contract no. DE-AC02-76SF00515 (synthesis and membrane devices), and the Air Force Office of Scientific Research (AFOSR) Hybrid Materials MURI under award no. FA9550-18-1-0480 (elasticity measurements and analysis).

\end{acknowledgements}

\section*{Supplementary Materials}
See supplementary materials for a more detailed discussion of strain distribution across the membrane upon loading with a spherical tip
and for notes on Weibull statistics.

\section*{Data availability statement}
The data that support the findings of this study are available from the corresponding author upon reasonable request.

%\bibliography{bibliography}% Produces the bibliography via BibTeX.
%\bibliographystyle{apsrev}

\providecommand{\noopsort}[1]{}\providecommand{\singleletter}[1]{#1}%

\begin{figure*}[t]
    \centering
    {\includegraphics[width=18 cm]{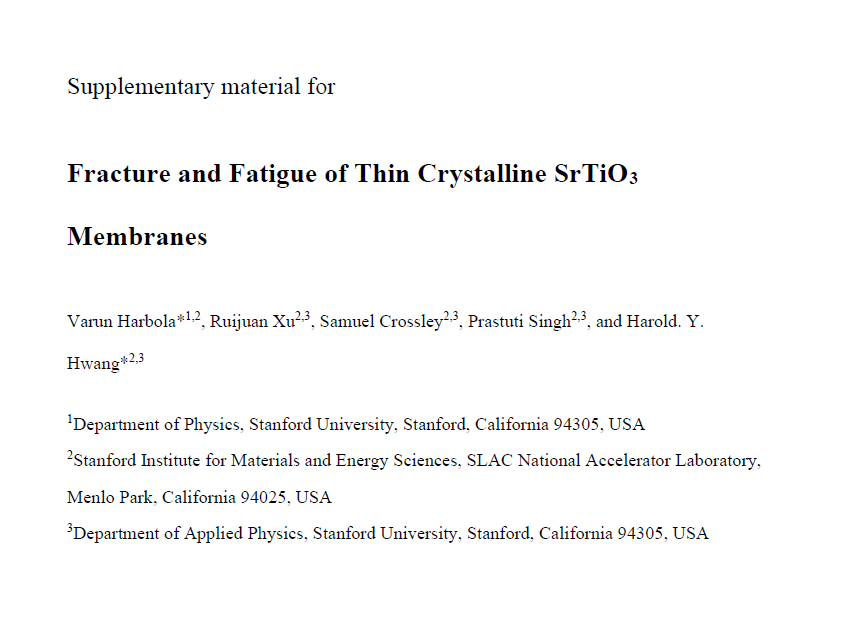}}
    
\end{figure*}

\begin{figure*}[t]
    \centering
    {\includegraphics[width=18 cm]{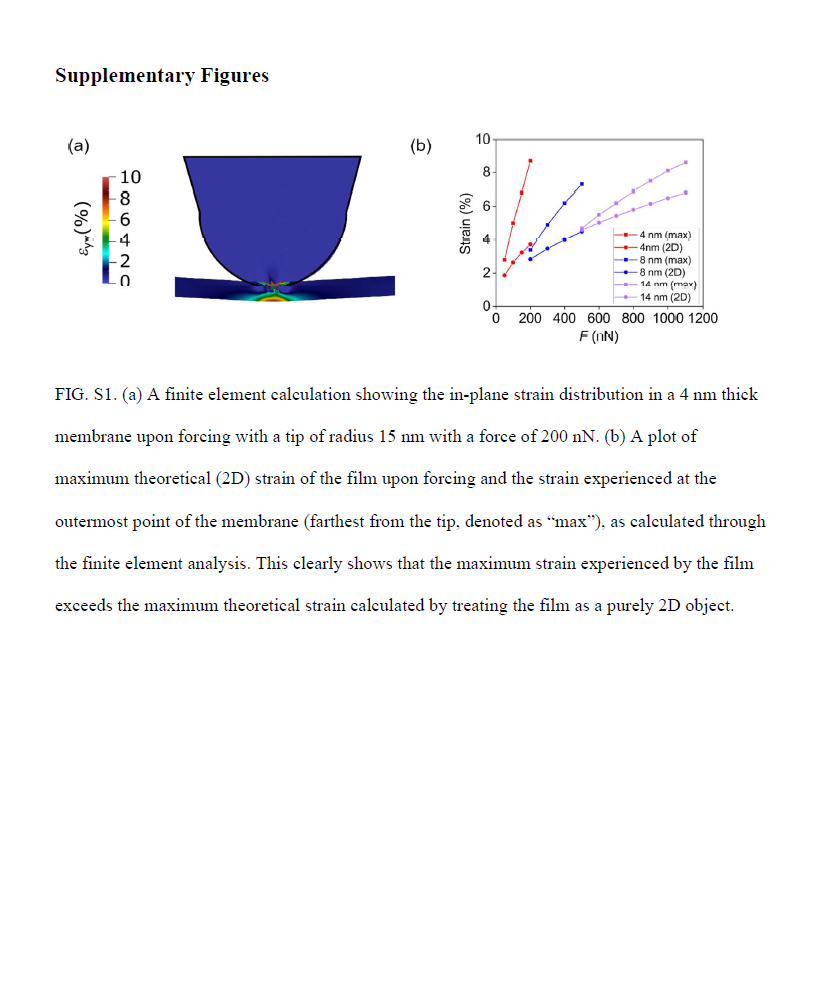}}
    
\end{figure*}

\begin{figure*}[t]
    \centering
    {\includegraphics[width=18 cm]{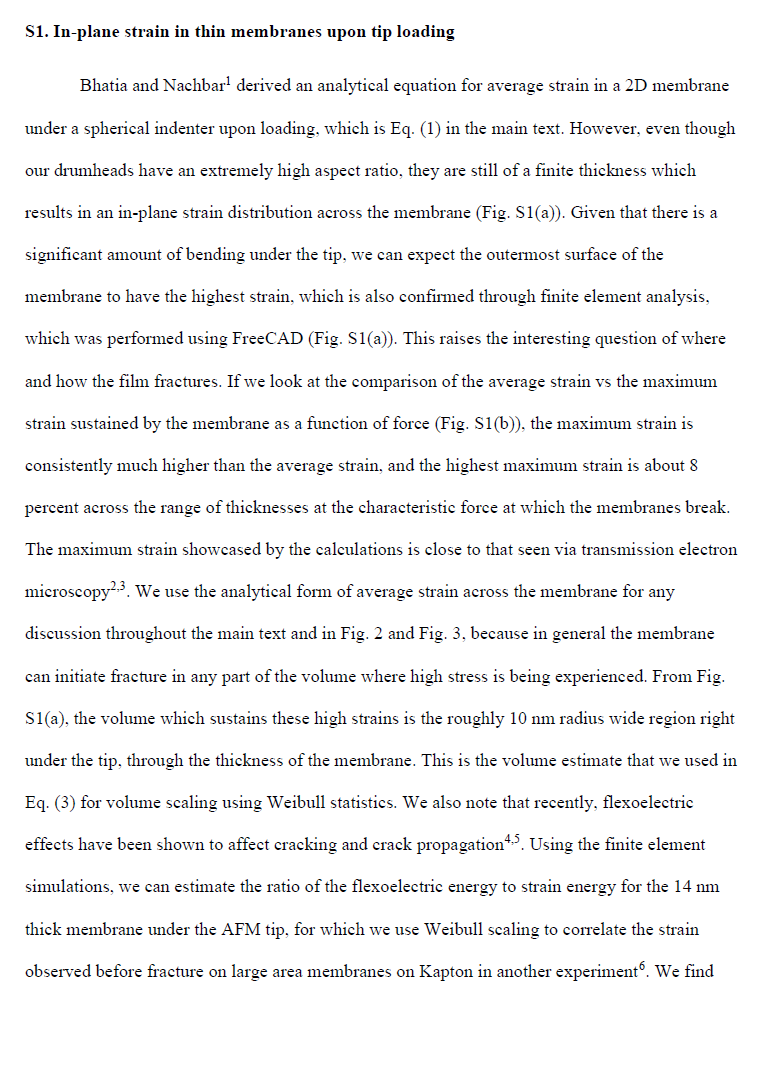}}
    
\end{figure*}

\begin{figure*}[t]
    \centering
    {\includegraphics[width=18 cm]{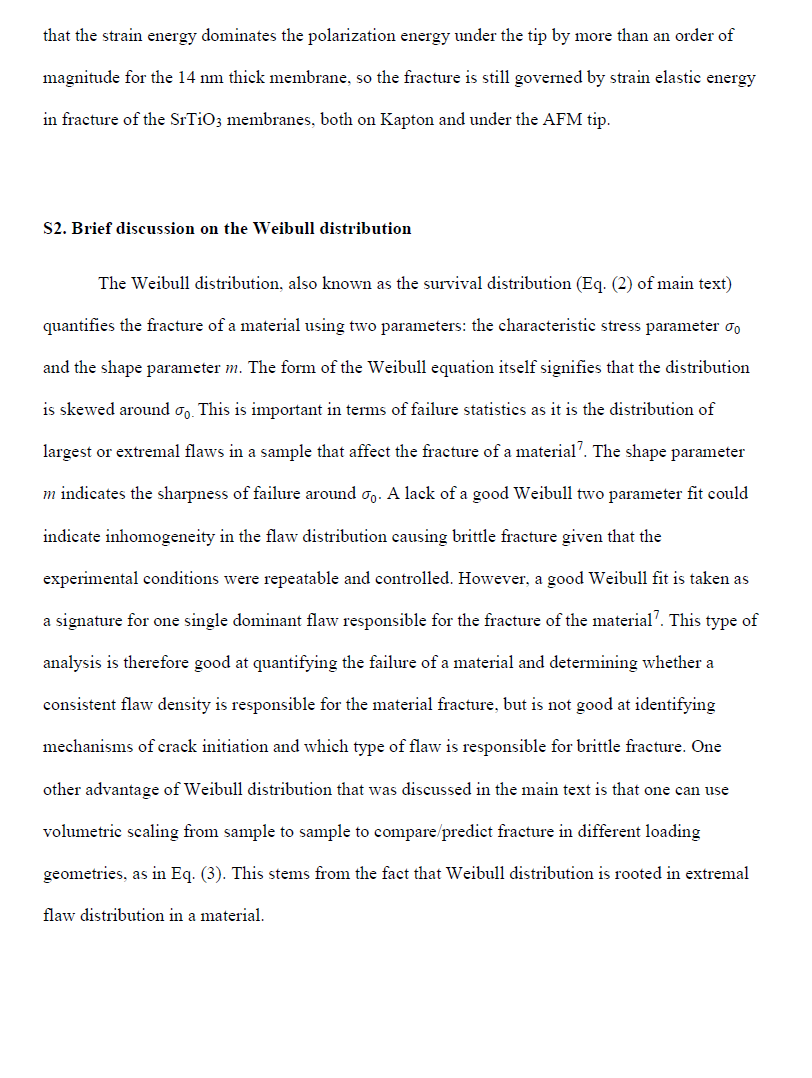}}
    
\end{figure*}

\begin{figure*}[t]
    \centering
    {\includegraphics[width=18 cm]{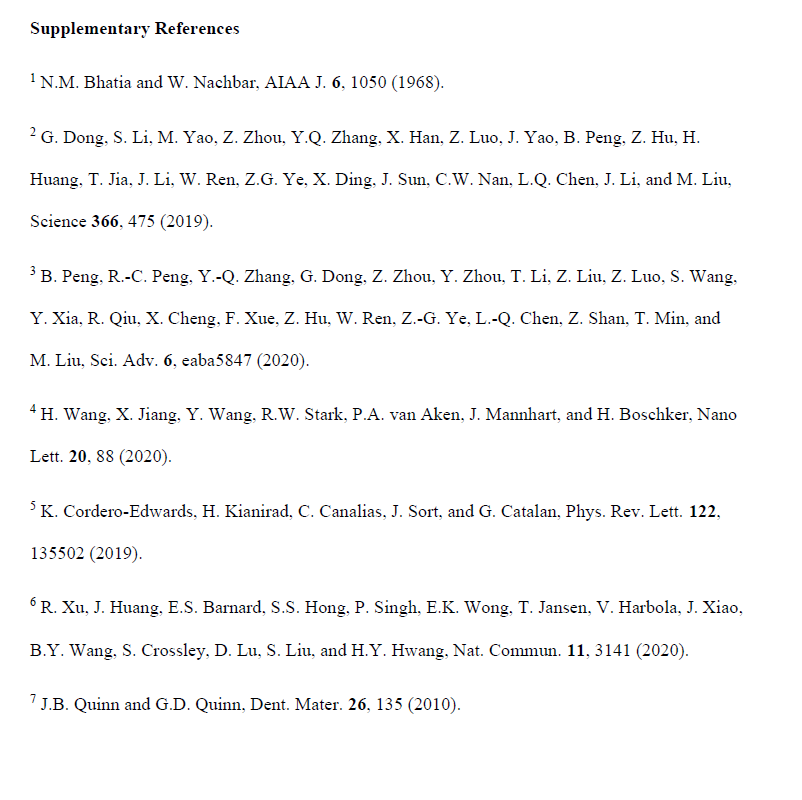}}
    
\end{figure*}

%\includepdf[]{Edited_supplementary_fracture_fatigue_srtio3.pdf}
\end{document}